\documentclass{emulateapj}

\shorttitle{Confirmation of the compactness of a galaxy with HST/WFC3}
\shortauthors{Szomoru et al.}

\begin{document}

\submitted{Accepted for publication in ApJ Letters}

\title{Confirmation of the compactness of a $z=1.91$ quiescent galaxy
  with Hubble Space Telescope's Wide Field Camera  3}

\author{Daniel Szomoru\altaffilmark{1}, Marijn Franx\altaffilmark{1}, 
    Pieter G. van Dokkum\altaffilmark{2}, Michele Trenti\altaffilmark{3}, Garth D. Illingworth\altaffilmark{4}, Ivo Labb\'e\altaffilmark{5}, Rychard J. Bouwens\altaffilmark{1,4}, Pascal A. Oesch\altaffilmark{6}, C. Marcella Carollo\altaffilmark{6}}

\altaffiltext{1}{Leiden Observatory, Leiden University, P.O.\ Box 9513, 2300 RA Leiden, The Netherlands.}
\altaffiltext{2}{Department of Astronomy, Yale University, New Haven, CT 06520-8101, USA.}
\altaffiltext{3}{University of Colorado, Center for Astrophysics and Space Astronomy, 389-UCB, Boulder, CO 80309, USA.}
\altaffiltext{4}{UCO/Lick Observatory, University of California, Santa Cruz, CA 95064, USA.}
\altaffiltext{5}{Carnegie Observatories, Pasadena, CA 91101, USA.}
\altaffiltext{6}{Institute for Astronomy, ETH Zurich, 8092 Zurich, Switzerland.}

\begin{abstract}
We present very deep Wide Field Camera 3 (WFC3)  photometry of a massive, compact galaxy
located in the Hubble Ultra Deep Field. 
This quiescent galaxy has a spectroscopic
redshift $z=1.91$ and has been identified as an extremely compact
galaxy by \cite{dad05}. We use new $H_{\mathrm{F160W}}$ imaging data
obtained with Hubble Space Telescope/WFC3 
to measure the deconvolved surface brightness
profile to $H \approx 28$ mag arcsec$^{-2}$. We find that the surface
brightness profile is well approximated by an $n=3.7$ S\'ersic profile.
Our deconvolved profile is constructed by a new technique which
corrects the best-fit S\'ersic profile with the residual of the fit to
the observed image. This allows for galaxy profiles which deviate from
a S\'ersic profile.  We determine the effective radius of this galaxy:
$r_e=0.42 \pm 0.14$ kpc in the observed $H_\mathrm{F160W}$-band. We
show that this result is robust to deviations from the S\'ersic model
used in the fit.  We test the sensitivity of our analysis to faint
``wings'' in the profile using simulated galaxy images consisting of a
bright compact component and a faint extended component. We find that
due to the combination of the WFC3 imaging depth and our method's
sensitivity to extended faint emission we can accurately trace the
intrinsic surface brightness profile, and that we can therefore
confidently rule out the existence of a faint extended envelope around
the observed galaxy down to our surface brightness limit. These
results confirm that the galaxy lies a factor $\sim 10$ off from the
local mass-size relation.

\end{abstract}


\section{Introduction}

A significant fraction of massive galaxies at $z\approx 2$ are
early-type galaxies containing quiescent stellar populations (e.g.,
\citealt{fra03}; \citealt{dad05}; \citealt{kri06}). These galaxies
must have formed very early in the universe's history and can
therefore provide important constraints on galaxy formation and
evolution models. Many of these quiescent galaxies have been found to
be extremely compact, with effective radii a factor $\sim 6$ smaller
than their low-z counterparts (e.g., \citealt{dad05}; \citealt{tru06};
\citealt{dok08}). This is quite puzzling, since these compact galaxies
are passively evolving and are therefore not expected to change
strongly in size or mass if they do not merge. We note that
\cite{man10} find some large massive quiescent galaxies at $z\sim
1.5$, showing that not all massive quiescent galaxies at high redshift
are compact.

Within the context of current-day models, galaxy mergers play an
important role in galaxy evolution (e.g., \citealt{whi91}). These
mergers may cause compact $z\sim 2$ galaxies to grow ``inside-out'',
i.e., the mergers would increase the size of the galaxies (e.g.,
\citealt{dok10}; \citealt{hop09b}). Whether the resulting size growth
is large enough, however, is uncertain (e.g., \citealt{bez09}).

Several authors have emphasised that there are several systematic
uncertainties that affect both radius and mass
determinations. First, effective radii may be underestimated due to
complex morphologies. Specifically, an extended low surface brightness
component could remain undetected due to low signal-to-noise ratio (S/N),
thereby lowering the observed size (e.g., \citealt{hop09a};
\citealt{man10}, but see \citealt{dok08}; \citealt{wel08}). Second,
mass-to-light gradients may result in a luminosity-weighted effective
radius that is smaller than the mass-weighted effective radius (e.g.,
\citealt{hop09a}; \citealt{hop09b}). Such gradients arise in certain
models for the formation of massive ellipticals (e.g.,
\citealt{rob06}; \citealt{naa07}). Finally, the inferred stellar masses
may be affected by incorrect assumptions regarding the initial mass
function (IMF) and stellar evolution models (e.g., \citealt{muz09},
and references therein).

In this Letter we use new very deep near-infrared (NIR) imaging data
from the Hubble Space Telescope's Wide Field Camera 3 (HST/WFC3) to
investigate the possibility of size underestimation due to the lack of
S/N. We examine the possibility of a ``hidden'' faint extended
component being present in $z\approx 2$ compact quiescent galaxies,
focusing on the most massive quiescent galaxy in the Hubble Ultra Deep
Field (HUDF) (Beckwith et al 2006), which has previously been studied by \citet{dad05}. We
adopt the following values for cosmological parameters: $H_0 = 70$ km
s$^{-1}$ Mpc$^{-1}$, $\Omega_m = 0.3$, and $\Omega_\Lambda = 0.7$. All
stellar masses are derived assuming a Kroupa IMF \citep{kro01}. All
effective radii are circularized, unless noted otherwise.

\section{Observations and sample}

\begin{figure}[h]
\plotone{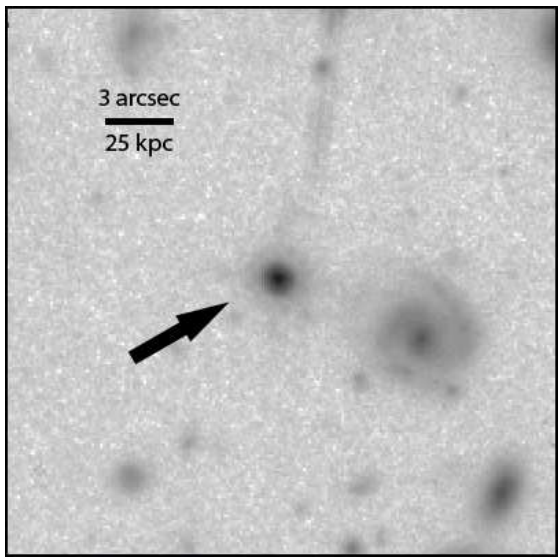}
\caption{The WFC3 $H_{\mathrm{F160W}}$-band image of the galaxy. It is
  well-separated from its nearest neighbors.}\label{fig:galim}
\end{figure}

Our study utilises new WFC3/IR $H_{\mathrm{F160W}}$-band imaging data
taken within the HUDF. These data are part of the first of the three
ultra-deep pointings which will be completed over the next year as
part of the HUDF09 HST Treasury program (GO11563). The current WFC3
imaging consists of 78600 seconds of exposure time in the
$H_{\mathrm{F160W}}$ band, leading to a limiting magnitude of
28.8. The post-spread function (PSF) 
FWHM is $\sim 0.16$ arcsec. Details of the data
reduction can be found in \cite{bou10}.

Since the WFC3 data does not cover the complete HUDF, most of the
compact massive $z\approx 2$ galaxies from e.g. \cite{dad05} and
\cite{cim08} fall outside of the observed area. From the compact
$z\approx 2$ galaxies inside the WFC3 HUDF image area we select the
most massive one, located at
$\alpha=3\mathbf{\,:\,}32\mathbf{\,:\,}38.12$,
$\delta=-27\mathbf{\,:\,}47\mathbf{\,:\,}49.63$. This galaxy has a
spectroscopic redshift $z=1.91$ \citep{dad05}, stellar mass $M_* =
0.56\times 10^{11} M_\odot$ (\citealt{wuy08}; F\"orster Schreiber in
preparation), and effective radius $r_e < 1$ kpc in the observed $z$
band (\citealt{dad05}; \citealt{cim08}). It was identified by
\cite{dad05} as passively evolving based on the $BzK$ criterion. A
summary of the galaxy's structural parameters is given in
Table~\ref{tab:prop}. An image of the galaxy is shown in
Figure~\ref{fig:galim}. It is sufficiently separated from its
neighbors to prevent contamination of its surface brightness profile.

\section{Fitting and size}\label{sec:psf}

\begin{deluxetable*}{l c c l c c}
\tablewidth{0pt}
\tablecaption{Structural parameters\label{tab:prop}}
\tablehead{
  \colhead{Source} & \colhead{$n$} & \colhead{$r_e$ (kpc)} & \colhead{$b/a$} & \colhead{$M_*$ ($10^{11} M_\odot$)} & \colhead{$H_\mathrm{F160W}^{tot}$ (AB)} }
\startdata
This Letter & $3.7 \pm 0.38$ & $0.42 \pm 0.14$ & $0.70$ & \nodata & $22.15 \pm 0.067$\\
Previous work & $4.7 \pm 0.6^1$ & $0.79 \pm 0.08^1$ & $0.74^1$ & $0.56^2$ & $22.12 \pm 0.03^2$ \\
\enddata
\tablenotetext{1}{\cite{dad05}, measured in the $z_\mathrm{F850LP}$ band}
\tablenotetext{2}{\cite{wuy08}}
\end{deluxetable*}

\begin{figure*}[t]
\plotone{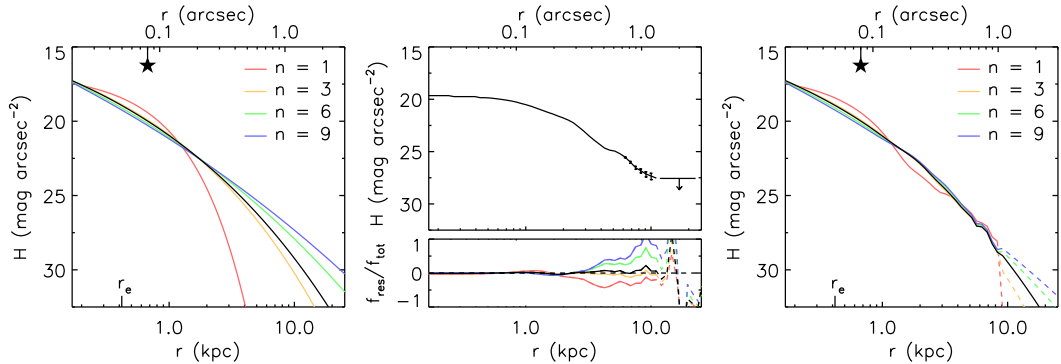}
\caption{Our method of correcting the observed surface brightness
  profile for the effects of the HST WFC3 PSF and incorrect profile
  modeling. In the left panel the best-fitting S\'ersic models,
  derived from a two-dimensional fit using a star as the PSF, are shown for
  different values of n. The black curve indicates the free-$n$ fit,
  with $n=3.7$. The profiles show large variations. In the top center
  panel the observed profile is shown. The residual fluxes from the
  S\'ersic fits are shown in the bottom panel as a fraction of the
  observed flux. In the right panel the profiles derived using our
  ``residual-correction'' method are shown. At large radii, where
  uncertainties in the sky determination become significant, the
  profile is extrapolated. This is indicated by dashed curves. The
  residual-corrected profiles are much more robust to modeling errors
  than the uncorrected profiles. The derived effective radius is
  indicated on the bottom x-axis, the PSF size (HWHM) is indicated by
  the star symbol on the top x-axis. The solid horizontal line in the
  middle panel indicates the $3\sigma$ sky noise level.  As can be
  seen, the surface brightness profile can be robustly measured to a
  surface brightness of 28 mag arcsec$^{-2}$.}\label{fig:profile}
\end{figure*}

We use the GALFIT package \citep{peng02} to fit two-dimensional
\citet{ser68} model profiles convolved with the PSF to the observed
surface brightness distribution.  This is an essential step in
deriving the structure of the galaxy, as the FWHM of the PSF of the
WFC3 images is significant compared to the size of the galaxy.  We use
a PSF extracted from a nearby unsaturated star and base our masking
image on a SExtractor (\citealt{ber96}) segmentation map. We fit nine
different models with a fixed S\'ersic index ($n=1,2,...,9$), as well as
a model where $n$ is a free parameter.

The effective radii from the S\'ersic fits range between 0.42 and 0.48
for S\'ersic indices varying between $n=1$ and $n=9$, with the
free-$n$ fit producing a value of 0.43 kpc (at $n=3.7$).  The best-fit
S\'ersic profiles are shown in Figure~\ref{fig:profile}. Despite the
fact that the effective radii are rather similar, it is clear that the
derived profiles vary significantly with $n$.

There is no intrinsic reason why galaxies should have ``perfect''
S\'ersic profiles. Although locally the surface brightness profiles of
elliptical galaxies are well fitted by single S\'ersic profiles over a
large range of radii (e.g., \citealt{kor09}), the situation may
be different at high redshift.
In particular, 
if elliptical galaxies grow by an inside-out process (e.g.,
\citealt{hop09a}; \citealt{fel09}), the surface brightness profiles of
their progenitors may deviate from S\'ersic profiles. We therefore
developed a method to derive more robust intensity profiles, which
depend less on the S\'ersic $n$ parameter used for the fit.

Our approach is the following: for each S\'ersic fit, we calculate the
residual image, which is an image of the observed flux minus the
PSF-convolved model. We derive a profile of the residual flux measured
along circles centered on the galaxy.  We add this residual profile to
the deconvolved model S\'ersic profile. We note that the intrinsic
profile is deconvolved for PSF, but the residuals are not. This
procedure is similar to how the CLEAN deconvolution method employed in
radioastronomy handles residuals \citep{hog74}. We thus remove or add
flux at those radii where the model does not adequately describe the
data, making a first order correction for errors caused by the 
incorrect profile choice. 
For large radii, where (systematic) uncertainties in the sky
determination become significant, we extrapolate the
residual-corrected profile by using the uncorrected S\'ersic profile,
scaled to the residual-corrected profile at the transition radius.
These ``residual-corrected'' profiles are then integrated in order to
determine the true half-light radius, which we refer to as $r_{e,
  deconv}$. The residual-corrected profiles are shown in
Figure~\ref{fig:profile}. The structural parameters of the
best-fitting profile are given in Table~\ref{tab:prop}.

It is clear from Figure~\ref{fig:profile} that the residual-corrected
profiles are much less sensitive to the S\'ersic-$n$ value adopted for
the initial modeling, especially at radii beyond a few
kpc. Furthermore, deviations from the S\'ersic profile are taken into
account; as we show in Section 4, using the residual-corrected profile
we can trace the true surface brightness profile much more accurately
than using simple analytical S\'ersic fits. This is due to the fact
that the S/N of the faint emission at large radii is so low that the
fitting procedure ignores it, even though a lot of flux can originate
there. Thus the stability of the parameters derived from S\'ersic fits
is no guarantee for correctness. This is particularly relevant when
the galaxies have complex morphologies, such as in the case of a
bright, compact galaxy surrounded by a faint, extended envelope.

Uncertainties in $r_{e, deconv}$ and the total $H$-band magnitude are
estimated from the range in values obtained from the fixed-$n$
residual-corrected profiles. The errors given in Table~\ref{tab:prop}
are the rms errors of the best-fit parameters from all of the fits,
and give an indication of the systematic errors due to differences
between the observed surface brightness profile and the S\'ersic
models used in the fitting procedure. The uncertainty in $n$ is
estimated using simulations: we add random sky noise to the observed
galaxy image. This is repeated several times, resulting in a number of
images, on each of which we perform the fitting procedure described
above. The uncertainty given in Table~\ref{tab:prop} is two times the
rms error of the best-fit parameter from all of the fits.

Our results are the following: the galaxy is best fit by a S\'ersic
profile with $n=3.7$. Using the residual-corrected profile we find
that the effective radius of the galaxy is $r_{e, deconv} = 0.050$
arcsec, which corresponds to $r_{e, deconv} = 0.42$ kpc. If we fix the
S\'ersic index to a constant value, the inferred size does not vary
substantially: $r_{e,deconv}$ varies from 0.31 kpc for
$n=9$ to 0.51 kpc for $n=1$. Thus, the deviations from the
best-fitting profile are $< 20\%$. Our size estimate
is therefore reasonably robust to deviations from the model profile.

We have investigated the influence of PSF uncertainties; if we use
PSFs extracted from other stars in the field we find variations in
$r_{e, deconv}$ of $< 10\%$.  We have used the Tiny Tim
software package\footnote{\tt http://www.stsci.edu/software/tinytim}
to investigate the spatial dependence of the PSF independently.  We
find that the derived effective radius changes very little with the
position of the reference star used, with a maximum of 10\% in
opposite corners of the field.  The difference in effective radius due
to the distance between the reference star and the galaxy is less
than 1\%.  We therefore conclude that PSF errors do not present a
significant problem in our analysis.

\section{Low surface brightness sensitivity}

\begin{figure*}
\plotone{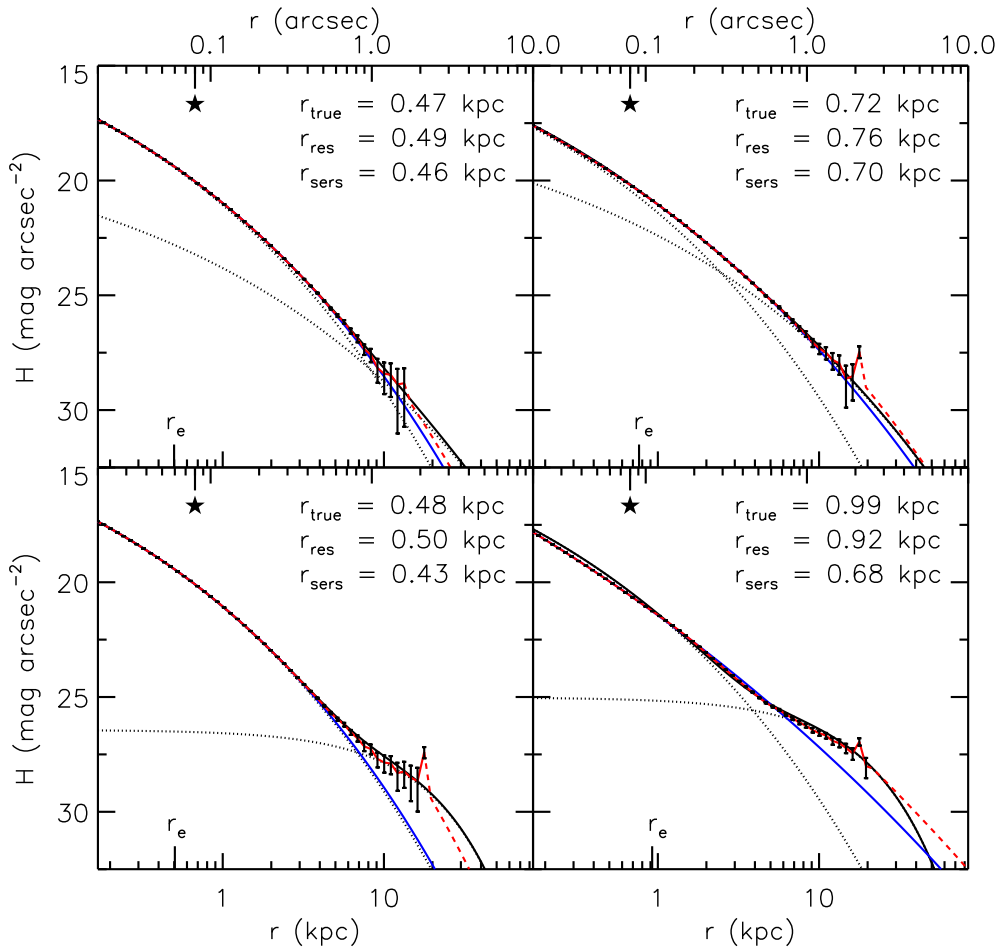}
\caption{Residual-corrected fits to simulated galaxy images (red
  curves). {\it Top:} compact $n=4$ profile and extended $n=4$
  profile, with flux ratios $10:1$ ({\it left}) and $2:1$ ({\it
    right}). {\it Bottom:} compact $n=4$ profile and extended $n=1$
  profile, with the same flux ratios as in the top panels. At large
  radii, where uncertainties in the sky determination become
  significant, the profile is extrapolated. This is indicated by
  dashed curves. The solid black curves indicate the total intrinsic
  surface brightness profiles, the dotted black curves indicate the
  individual components that make up these profiles. The best-fit
  uncorrected S\'ersic profiles are shown as blue curves. These
  deviate strongly from the true profiles at large radii. The
  residual-corrected profiles (red curves) follow the intrinsic
  profiles extremely well, demonstrating that our method recovers the
  intrinsic profiles accurately; the derived effective radii,
  indicated on the bottom x-axes, are within 10\% of the true
  effective radii. The PSF size (HWHM) is indicated by the star symbols
  on the top x-axes.}\label{fig:sim}
\end{figure*}

We now determine whether faint extended emission would be detected
using our data. To this end we construct several simulated galaxy
images which consist of two components; a compact component, described
by a $n=4$ S\'ersic profile with an effective radius roughly equal to
the observed galaxy (see Table~\ref{tab:prop}), and an extended
component, described by a S\'ersic profile with either $n=4$,
$r_e\approx 3.5$ kpc or $n=1$, $r_e\approx 15$ kpc. The extended
component has a flux that is either 10\% or 50\% of the compact
component's flux. The compact component's flux is chosen such that the
total flux of the two components is equal to the observed galaxy's
total flux.  The images are convolved with the PSF, and sky and
readout noise are added.  The images are then fit with a single
S\'ersic profile using GALFIT, and a residual-corrected profile is
constructed. By comparing the half-light radii obtained in this way to
the intrinsic half-light radii we can quantify the sensitivity of our
data to low surface brightness components.

The results of our simulated galaxy fits are shown in
Figure~\ref{fig:sim}. The residual-corrected profiles closely follow
the intrinsic profiles. The effective radii derived from the
residual-corrected profiles are very close to the intrinsic effective
radii: in three of the cases the difference is less than $5\%$. For
the $n=1$ extended component with a total flux equal to half of the
compact component's flux the inferred radius is 10\% smaller than the
intrinsic radius, comparable to the systematic error due to modeling
uncertainties (see Section~\ref{sec:psf}).  We also tested $n=4$ and
$n=1$ models with effective radii of several kpc for the $n=1$
extended component: these models are so well approximated by S\'ersic
models with higher values of $n$ ($>4$) that normal S\'ersic
profile fitting immediately retrieves the correct effective radii.

In conclusion, our method used on these deep data is sensitive to a
faint extended component down to a surface brightness of $H\approx 28$
mag arcsec$^{-2}$, and using our method we retrieve effective radii
that are within $1\sigma$ of the true value. We note that the
effective radii obtained using the conventional method are, in most
cases, very close to the intrinsic effective radii. However, the
surface brightness profiles obtained in this way clearly deviate from
the intrinsic profiles.

\section{Discussion}

\begin{figure*}
\plotone{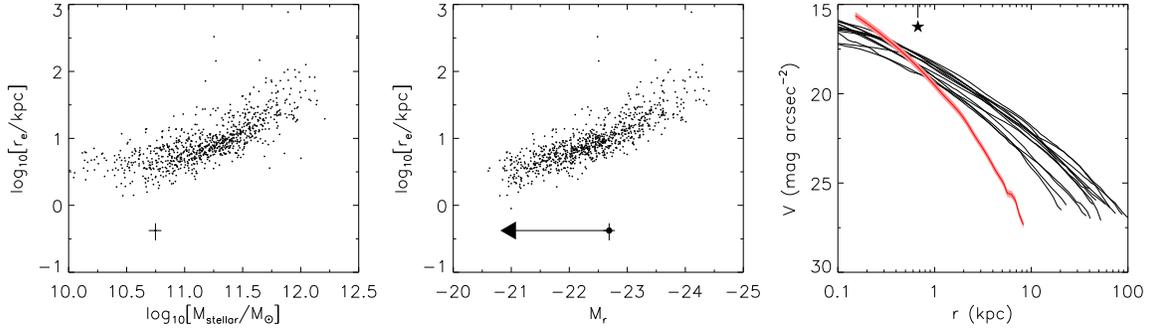}

\caption{Relations between size and stellar mass ({\it left}) and size
  and rest-frame $r$-band luminosity ({\it middle}) for a sample of
  low-redshift galaxies, taken from \cite{guo09}. The large symbol
  with error bars indicates the position of our galaxy. Low-redshift
  galaxies are much larger at similar stellar masses and
  luminosities. The arrow in the middle plot indicates the change in
  luminosity due to passive evolution to $z=0$. The size of the galaxy
  is smaller than the local equivalents by a factor of 10.  {\it
    Right:} comparison of best-fit residual-corrected rest-frame
  V-band surface brightness profile (red curve) to elliptical galaxies in the
  Virgo cluster, from \cite{kor09} (black curves). Virgo galaxies are plotted
  with
  $M_*/M_\odot > 10^{11}$.
  The observed high-$z$ surface brightness profile has been
  corrected for cosmological surface dimming and passive M/L evolution
  from $z=1.91$ to $z=0$ (see text). Assuming the galaxy has a mass $>
  10^{11} M_\odot$ at $z=0$ its profile at large radii will evolve
  very strongly over the next 10 Gyr. The central surface brightness
  profile, on the other hand, shows much less evolution between
  $z\approx 2$ and $z=0$.}\label{fig:rel}

\end{figure*}

We have found that the galaxy under consideration is indeed remarkably
small.  We have fitted a S\'ersic model to the observed flux
distribution, and corrected the profile for the observed deviations.
We have measured the galaxy's half light radius: $r_{e, deconv} = 0.42
\pm 0.14$ kpc. This result is robust to changes in the imposed
S\'ersic profile. As a check of our data's sensitivity to a low
surface brightness component we have constructed simulated galaxy
images which include a faint extended component. We can reproduce the
effective radii to 10\% using our technique.

A possible cause for concern is that the galaxy might deviate strongly
from a S\'ersic profile.  We have incorporated the residuals in our
fit to compensate for such errors, and we note that the residuals from
our best S\'ersic model fit are quite low ($<10\%$). This implies that
our model profile is close to the real profile. This, and the fact
that varying $n$ has little influence on the derived half-light
radius, suggests that our results are not strongly affected by this
source of error.

Thus, our findings indicate that the small effective radius that has
been found is not due to oversimplified modeling or a lack of S/N, and
gives additional evidence that a strong evolution in size occurs from
$z\approx 2$ to $z=0$. It should be noted that our derived effective
radius is $1.6$ times smaller than the radii derived by \cite{dad05}
in the $i$ and $z$ bands.  When we repeat our analysis on the ACS
$z$-band data we obtain a slightly different value, $r_{e, deconv}
\approx 0.65$ kpc (uncircularized), closer to the deep $H$-band
imaging, and somewhat smaller than the value derived by \cite{dad05}
(but consistent within the errors).  Hence all bands indicate a very
small size.

Figure~\ref{fig:rel} illustrates the difference in size and mass
between our galaxy and the $z=0$ elliptical population; plotted in the
first two panels are the compact galaxy we have studied and a sample
of low-redshift central galaxies from groups and clusters in the Sloan
Digital Sky Survey, analyzed by \cite{guo09}. The compact $z\approx 2$
galaxy lies far off from the $z=0$ mass-size relation. The middle
panel shows the galaxy on the mass-luminosity relation.  We estimated
the luminosity evolution of the compact galaxy from $z=1.91$ to $z=0$
in two ways: we first used the rest-frame $B-I$ color difference
between low and high redshift to estimate the difference in
mass-to-light ratio. Second we used the Fundamental Plane to estimate
the evolution from $z=0$ to $z=1$ from \cite{wel05}, and used the
average evolution of the mass-to-light ratios of early-types in the
CDFS at $z=1$ and the $z=1.91$ galaxy, both from F\"orster Schreiber et al. (in preparation).  The
resulting evolution is 1.8-2.2 magnitudes.  As a result, the galaxy
still lies off from the size-magnitude relation after correcting for
evolution.

In the third panel of Figure~\ref{fig:rel} we compare the surface
brightness profile of this galaxy to those of elliptical galaxies in
the Virgo cluster. The profile shown has been corrected for
cosmological surface brightness dimming and passive luminosity
evolution from $z=1.91$ to $z=0$. The total correction is $-3.5 + 2
\approx -1.5$ magnitudes. Even though the galaxy has an average
density $>100$ times larger than the average $z=0$ elliptical of the
same mass, its surface brightness profile in the central kpc is
actually rather similar to those of the most massive galaxies at $z=0$
- the average density measured at fixed physical radius is not that
different.  This is consistent with results obtained by other authors
(e.g., \citealt{bez09}; \citealt{hop09a}; \citealt{fel09};
\citealt{dok10}).  Thus, the main difference between $z=0$ and this
$z\approx 2$ galaxy is at larger radii where the $z\approx 2$ galaxy
has much lower surface brightness. Such a result could be explained by
inside-out growth.

We also  note that there may be significant errors in the mass
determination of $z\approx 2$ compact galaxies, due to e.g.\ incorrect
assumptions about the IMF. Changes in the low mass end of the IMF
affect both the masses of the high redshift and low redshift galaxies,
and are nearly irrelevant. However, changes in the slope of the IMF
will affect the derived passive evolution between $z=2$ and $z=0$, and
will increase or decrease the size evolution. Changes in the IMF could
thus have important consequences for evolution.  Future deep NIR
spectroscopic data should provide direct information on the kinematics
of these objects and will allow us to confirm their high masses (see
e.g., \citealt{dok09}).

Finally, it will be interesting to obtain similar deep data on other
compact massive galaxies, so that their profiles can be analyzed to
the same surface brightness limit. We note that stacking can also lead
to a great increase in imaging depth; e.g., \cite{cas09},
\cite{dok08}, and \cite{wel08} stack samples of compact galaxies and
obtain very good constraints on their average surface brightness
profile. However, with the new WFC3 data available in the coming years
many more compact massive galaxies can be studied on an individual
basis.

We acknowledge  support from NASA grant HST-GO-11563
and ERC grant 227749.

\end{document}